# Photophysics of pentacene-doped picene thin films

Tullio Toccoli[1*], Paolo Bettotti[3], Antonio Cassinese[4], Stefano Gottardi[1], Yoshihiro Kubozono[5], Maria A. Loi[2], Marianna Manca[2], Roberto Verucchi[1]

[1]*IMEM–CNR, Institute of Materials for Electronics and Magnetism, Trento Unit, Via alla Cascata 56/C – Povo, 38123 Trento, Italy*

[2]*Photophysics and OptoElectronics, Zernike Institute for Advanced Materials, University of Groningen, Nijenborgh 4, Groningen 9747 AG, The Netherlands*

[3]*Nanoscience Laboratory, Department of Physics, University of Trento, Italy.*

[4]*SPIN-CNR and Physics Department, University of Naples, P.le Tecchio 80, I-80125 Naples, Italy*

[5]*Research Institute for Interdisciplinary Science, Okayama University, Okayama 700-8530, Japan.*

\* ***Author to whom correspondence****: Tullio Toccoli; Adress: IMEM–CNR, Institute of Materials for Electronics and Magnetism, Trento Unit, Via alla Cascata 56/C – Povo, 38123 Trento, Italy; E-Mail: tullio.toccoli@unitn.it;*

**Abstract**

Here were report a study of picene nano-cristalline thin films doped with pentacene molecules. The thin films were grown by supersonic molecular beam deposition with a doping concentration that ranges between less than one molecules of pentacene every $10^4$ picene molecules up to about one molecule of pentacene every $10^2$ of picene. Morphology and opto-electronic properties of the films were studied as a function of the concentration of dopants. The optical response of the picene films, characterized by absorption, steady-state and time-resolved photoluminescence measurements, changes dramatically after the doping with pentacene. An efficient energy transfer from the picene host matrix to the pentacene guest molecules was observed giving rise to an intense photoluminescence coming out from pentacene. This efficient mechanism opens the possibility to exploit applications where the excitonic states of the guest component, pentacene, are of major interest such as MASER. The observed mechanism could also serve as prototypical system for the study of the photophysics of host guest systems based on different phenacenes and acenes.

## 1. Introduction

Acenes and phenacenes are two different series of polycyclic aromatic hydrocarbon (PAH) compounds of fused benzene ring arranged into linear and zigzag sequence, respectively.[1] Picene and pentacene are isomeric forms that show contrasting electronic and optical properties because of the different arrangement of the rings that modifies the energetic levels of the conjugation system.[1,2,3] Molecules of the acenes series have been known for a long time to exhibit outstanding transport properties, so that they became model systems for organic electronics.[1] The ultrapure single crystals of anthracene,[4] tetracene,[5,6] as well as pentacene demonstrated semiconducting properties with high carrier mobility values comparable to those of amorphous inorganic materials, (up to 40cm$^2$/Vs for pentacene[7] and rubrene[8]). Some of these molecules also displayed ambipolar transport with appreciable charge carrier mobilities.[9] Thin films of acenes has been grown with different techniques including evaporation,[10] supersonic molecular beam deposition,[11] and even by solution processing through molecular modification.[12] Besides their superb transport properties, members of the acene family were recently investigated also for the singlet exciton fission due to the relative positions of their singlet and triplet excited states.[13,14,15] This phenomenon is currently gaining interest not only in the photovoltaic community to enhance the efficiency of organic solar cells but also for the realization of room temperature solid state MASER as it was recently reported for the case of pentacene included in p-terphenyl crystals.[16,17] For these reasons, we present here a study of a promising host-guest system based on picene and pentacene. The choice of picene as host molecules was made for its stability in ambient conditions and for its electronic, optical and electrical characteristic.[18] Pentacene is a well-known molecule for organic electronics[19] that is believed to have a good affinity with picene due to the structural similarities and optical complementarities. We show in the following how the host-guest system made by picene and pentacene could be interesting both for the possible use in fabrication of optoelectronic and microwave devices and for the study of the molecular properties that are otherwise not accessible in isolated molecular form.

In this article, we demonstrate a strong enhancement of the photoluminescence (PL) intensity of pentacene once embedded into a picene matrix. The PL increase, is shown to originate from a very fast energy transfer process between picene and pentacene that seems to be of Förster type. The probably distortion of pentacene molecules embedded in the picene matrix could also favor its emission. In fact, the doping of a picene matrix allows to effectively access the energy levels of the isolated pentacene molecules, overcoming the problem of its poor solubility and low emission intensity. The high efficiency of the energy transfer could also enable the access to the optical levels of pentacene (i.e. triplet state) without the need of a strong optical excitation that could damage the material. In the following, we correlate the material optoelectronic properties with, the growth conditions and film morphology by means of atomic force microscopy (AFM) analysis, ultraviolet photoelectron spectroscopy (UPS), absorbance, PL and time resolved photoluminescence spectroscopy.

## 2. Experimental Section

Thin Film Growth: Pentacene doped picene thin films were grown by supersonic molecular beam deposition (SuMBD) in ultra-high vacuum (UHV). A detailed description of the SuMBD setup can be found elsewhere.[20] Two separate crucibles were loaded inside the SuMBD source: one with picene powder acquired from (Asahi Kasei, 99%) and one with pentacene powder (Sigma Aldrich, 99.995% pure). The temperature gradient inside the supersonic source can be controlled by acting on two separate heating stages positioned at the nozzle and around the body of the source. The relative concentration of the compounds in the molecular beam can be regulated within a certain range and precision (see hereafter) by controlling the temperature gradient inside the source and by considering the different vapor pressure of the two materials. Additionally, the system is provided with a time of flight mass spectrometer that permits to measure the mass spectra and the flux of the molecules in the beam.[20] The molecular beam was seeded in He carrier gas and characterized by a kinetic energy per molecule of about 7 eV. All depositions were performed in UHV (background pressure $10^{-9}$ mbar) on thermal silicon oxide (50 nm thick) on silicon or on quartz substrates kept at room temperature, with a growth rate of about 0.3 monolayer/minute. In this study we fabricated three sets of samples with different pentacene concentration. Set 1 where the pentacene concentration was estimated to be lower than one molecule every $10^4$ molecules of picene. Set 2 where the pentacene concentration was 10 times higher respect to set 1 and set 3 with a pentacene concentration of about one molecule every 100 molecules of picene. To determine the concentration we assumed that the sticking coefficients of the two molecules on the substrate are similar and the final stoichiometry of the films is the same of that present in the supersonic molecular beam. Picene and pentacene are stereoisomers and their peaks in the mass spectrum collected with our spectrometer are in the same position; for these reasons, we do not have a direct access to their relative concentration in the supersonic beam. To do this we worked with only one crucible filled alternatively with one of the two materials and keeping the same conditions used for growing the thin film. The value for the concentrations was calculated from the ratio between the area underlying the picene and pentacene peaks in the mass spectra.

Thin Film Characterization: The morphology of the films was characterized by AFM in air, using a Smena SFC050 scanning head by NTMDT. Measurements were carried out in semi-contact mode AFM (using NSG11 silicon cantilevers by NTMDT). The detailed analysis of AFM data was performed with the WSxM software (version 5.0 Develop 1.1, Nanotec electronica).[21]

Organic thin films have been characterized ex-situ by UPS, using the He II photon emission at 40.82 eV, produced by a Helium discharge lamp. The hemispherical electron energy analyzer (VSW HA100, interfaced with a PSP power supply unit) had a total energy resolution of 0.86 eV. The binding energy (BE) position for valence band was calibrated by using as reference the Au Fermi level (0 eV). All analyzed films were deposited on 50 nm $SiO_2$ layer on Si, leading to strong charging effects during UPS. We used this thick $SiO_2$ layer to avoid any substrate influence on observed electronic properties to properly compare with all other analysis.

The UV−Vis−NIR spectra were collected in a Varian Cary 5000 spectrometer in the range 1000-200 nm with a scan rate of 150 nm/min and a bandwidth of 2 nm in a double beam configuration.

Photoluminescence measurements were performed on a Cary-ECPLISE fluorometer in the spectral range 350-600 nm, the excitation wavelength was set to 350 nm. Both excitation and emission slit widths were 10 μm and a scan rate of 120 nm/min was used. The instrument automatic excitation filter was used to reduce the effect of higher diffraction orders. Time resolved PL measurements were performed by exciting the samples with the second harmonic (380 nm) of a mode locked Ti:Sapphire femtosecond laser (Coherent), tunable in the range ~ 720-980 nm, characterized by pulses of ~150 fs and a repetition frequency of ~ 76 MHz. The PL spectra were detected with a monochromator coupled with an Image EM CCD camera (Hamamatsu Photonics). The spectra were corrected for the spectral response of the set up. Time resolved traces were recorded with a streak camera (Hamamatsu Photonics) working in synchroscan mode.

## 3. Result and Discussion

The picene-pentacene host-guest system (pentacene inside a picene matrix) is an interesting reference system for the study of small-molecules nanocrystalline compounds.

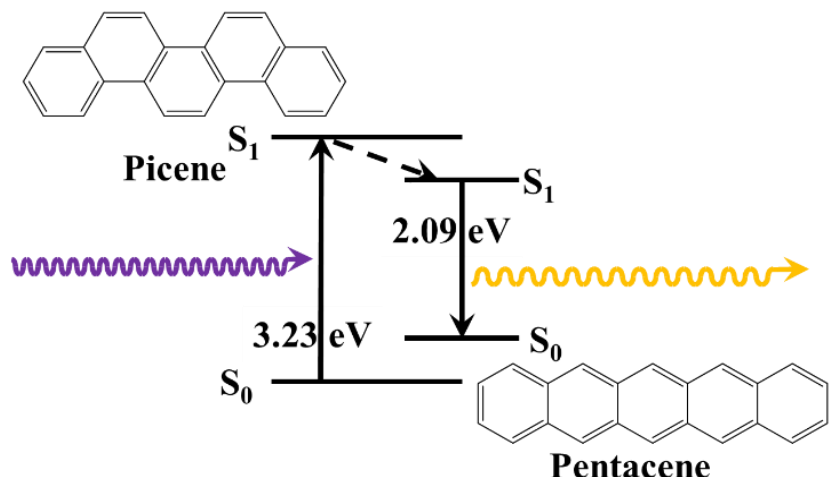

**Figure 1.** Scheme of the excitonic levels of picene and of pentacene and their molecular structure. We indicate the singlet energy level of picene in thin film (about 3.3 eV) and the one of molecular pentacene (about 2.1 eV). The process of the energy transfer from the host (picene) to the guest (pentacene) is represented: optical excitation (violet arrow), energy transfer (black dashed arrow) and radiative decay (orange arrow).

The singlet energy levels of these molecules (Figure 1) are well-positioned to expect efficient transfer of the excitation from the host to the guest. Both molecules in the thin film show a similar crystal structure[22,23] with similar interplane distance (about 1.4 nm for picene[22] and about 1.5 nm for pentacene[23]) and their molecular dimensions are very close (1.15 nm for picene and 1.22 nm for pentacene)[2] and also their thin films packing, it is possible for pentacene molecules to substitute picene molecules while keeping the same crystal packing and thus minimizing phase segregation in the thin film. We can expect, based on their crystalline structures and molecular dimensions, that in these conditions pentacene will be distorted by the crystal field of picene modifying its symmetry and consequentially some of its chemical and physical properties.

We know that in the thin film growth (in terms of nucleation, island formation and coalescence) these two molecules follow different pathways[22] giving as result a different morphology.[22,24] However, due to the low concentration of pentacene in the thin film, the growth and the final morphology are expected to be determined by the host component. This is confirmed by the comparison of thin films morphologies reported in Figure 2. Here we show the morphological characterization of a pure picene film (a) a mixed film of set 1 (b), and pure pentacene (c). All films were grown on thermal silicon oxide surface (50 nm thick). As we have reported previously[22] the early-stage morphology of pure picene films shows that the molecules nucleate in islands that develop in a 3D growth. The crystallites are several hundred of nanometers wide with sharp edges and very regular shape; the average height of the crystallites is about 70 nm.

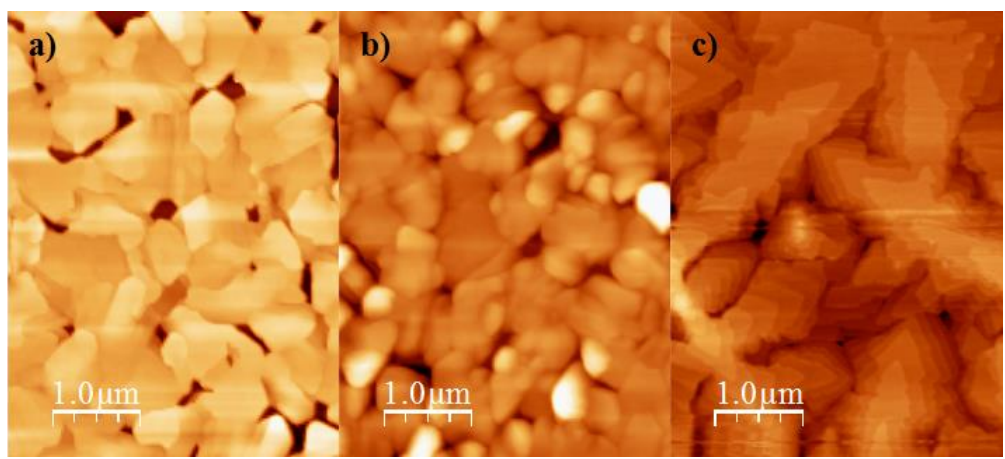

**Figure 2.** AFM micrograph: a) pure picene; b) set 1 samples; c) pure pentacene. b) In the mixed film no phase segregation is visible. The morphology is substantially the same of the pure picene film.

Doping picene with a concentration of pentacene $< 10^{-4}$ (set 1 samples) doesn't change significantly the thin film growth (Figure 2b), but the crystallites seem to be less regular (we observe that the typical sharp edges present for pure picene are not so evident anymore). The average height of the crystallite before reaching percolation is about 70 nm as for pure picene on silicon oxide.[22] From the AFM characterization we were not able to observe presence of crystallites displaying the typical pentacene growth (figure 2c) and thus we exclude the possibility of phase segregation.

Figure 3 reports the analysis of the valence band (using as photon source the He II emission at 40.82 eV) for pure pentacene and picene films, as well as picene film with two different pentacene doping concentrations (set 1 and set 3). The lineshape of the spectra shows differences with reported data in the literature.[25,26] These differences are related to the different ionization cross section of the exciting photon used in this work.[27,28,29] Nevertheless, the HOMO features for pentacene and picene pure films at about 0.9 and 2 eV can be easily identified, in agreement with previous results.[25,26] Other differences between the two molecules valence bands can be found in the 6-11 eV energy region. Comparing picene films with those doped by different amount of pentacene, apparently no significant energy shifts occur for its HOMO and main features. The overall lineshape is partially altered, with features becoming less defined and broader. By comparison with the valence band of pure pentacene, the observed changes could be attributed to the appearance of bands related to the guest molecules. At the higher concentration, the appearance of a broad peak at about 1 eV can be identified (see Figure 3 right), i.e. in the BE region of the pentacene HOMO.

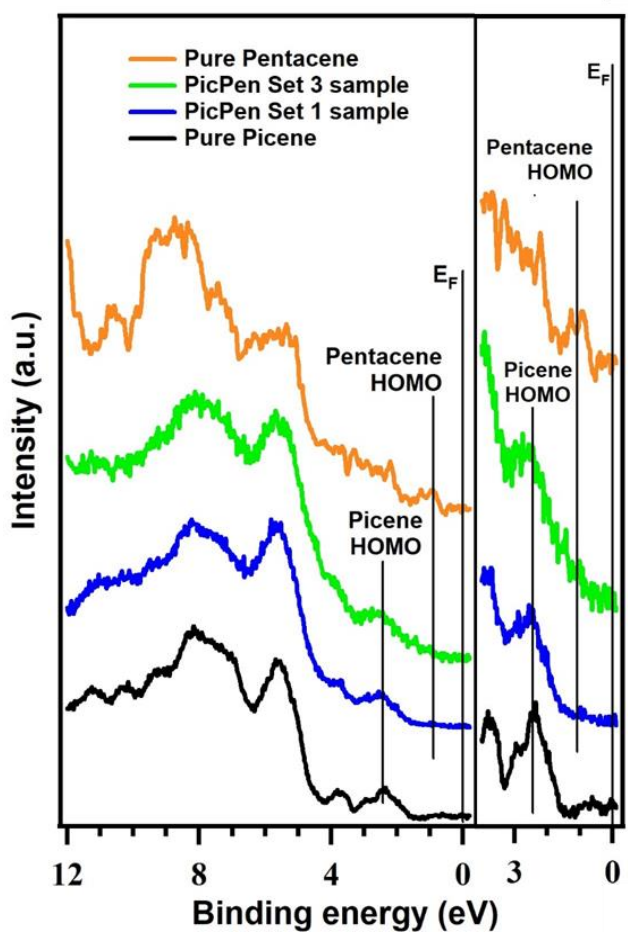

**Figure 3.** UPS (He II) valence band analysis of organic thin films of pure pentacene (orange curve), pure picene (black curve) and picene with different concentration of pentacene (blue, low; green, high). Films are deposited on 50nm SiO$_2$ layer on Si. Valence band are normalized in height, with markers for the picene and pentacene HOMO features, Fermi energy lev-el (EF).

This structure can be definitely attributed to the pentacene HOMO, while more detailed considerations are not possible due to the low intensity and large width of this peak. Being its energy position strictly similar, to that of pristine pentacene film, this suggests a simple superposition of the two valence bands with a Fermi levels alignment. We can conclude that the pentacene HOMO and

LUMO levels are inside the HOMO-LUMO gap of picene. The position of the levels of the two molecules suggests that an efficient energy transfer from picene to pentacene is possible. With this host-guest system it may be possible to access excited states of the pentacene molecule that are not easily accessible otherwise. This opens the possibility to use such host-guest system for several applications like lighting, photovoltaics[13] and solid-state MASER.[16]

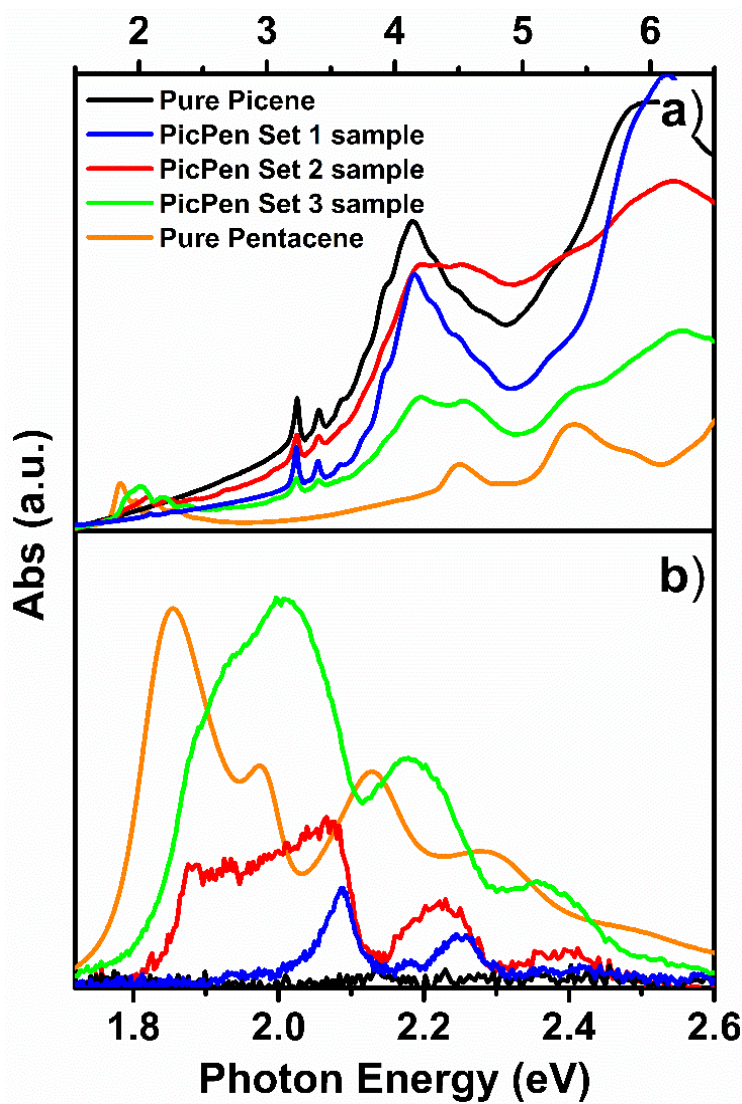

**Figure 4.** a) Absorbance spectra of the different set of samples compared with the one of pure picene (black) and pure pentacene (orange). Blue line set 1 sample; red line set 2 sample; green line set 3 sample. b) Detailed area of pentacene absorption area. Here we have subtracted manually a baseline to enhance the peaks view.

The doping of the films has a much greater impact on the optical properties than on morphology and molecular energy levels. The exciton dynamics of pure picene is strongly modified by the presence of pentacene and thus the photophysical properties of the host-guest system are strongly dependent on the pentacene concentration. In the following, we are going to show the result for the absorption, the PL emission and the time resolved PL. Figure 4a reports the absorption spectra for each set of the pentacene-doped picene and for the pure picene and pentacene thin films. In all the spectra (except for the pure pentacene) it is possible to observe a main spectral region (3.0-6.0 eV), which corresponds to the absorption of picene that is weakly affected by the presence of pentacene in the doped films. All thin films spectra display a relatively broad band at about 4.13 eV and 3 narrow peaks at 3.24, 3.41 and 3.58 eV (this last one is a shoulder of the broad band at 4.1 eV and is clearly visible only on pure picene and on set 1 samples). Similar features have been theoretically predicted by Rubio et al. using Bethe-Salpeter equation,[30] and experimentally confirmed by Xin et al. in the single crystal phase of picene.[31] For the set 3 samples, (green line) we can observe also the presence of shoulders at about 4.5 eV and at 5.4 eV. These bands are due to pentacene as we can see by comparison with its spectrum (orange line). At high energy the absorption spectra could be reasonably explained by the sum of the contribution from both picene and pentacene spectra.

This situation changes in the case of the spectral region between 1.5 eV and 2.7 eV where the picene does not show any absorbance (see figure 4b, where we re-ported a detailed view of this spectral area corrected for a baseline for clarity). Here we observe that the spectra cannot be ascribed to the simple overlap of the absorption of thin films of picene and pentacene, but the spectra show peaks in other positions (i.e. around 2.08 eV and 2.25 eV) that we have ascribed to absorbance of dispersed pentacene molecules inside the picene matrix. We observed that these peaks change their spectral position (red shift) and shape with increasing the pentacene concentration inside the film with a trend that seems to move toward the absorption of pure pentacene (compare green line and orange line).

In more detail: for set 1 samples (lower pentacene concentration) these bands are very sharp and localized at 2.09 eV and 2.26 eV. Both are assigned to the absorption of single pentacene molecules as reported in reference.[32,33] Here the pentacene molecules are so dispersed in the picene host matrix that there is no interaction between them and their absorption becomes very similar to that of low concentration pentacene in a frozen solution.[33]

Increasing the pentacene percentage, (red line) we note a change in the shape of the low energy band that is now very large and seems composed of several peaks. The first one clearly visible (red line) is located at about 1.89 eV while the last one located now at about 2.07 eV have probably the same origin of the low energy transition for the low pentacene concentration film (blue line). In the case of this intermediate pentacene concentration we observe also a red shift of the second band that are now located at 2.25 eV.

The red shift, the shape change and the appearance of the new peaks indicate that increasing the pentacene concentration the molecules interact with each other modifying their optical absorbance from single dispersed molecules towards molecular aggregates.[33] For example a J-like aggregation can determine the red-shift of the absorption.[34]

Despite the samples have been exposed to ambient condition for a long time and the optical measurements were performed in air, interestingly, there is no signature of pentacene-quinone[35] in the spectra that would result in absorption peaks at 2.90 eV and 3.11 eV. This observation indicates that the picene matrix can protect the pentacene molecules from (photo)oxidation. In this way, the probed optical characteristics of the guest molecules in this system come solely from the pure pentacene, not from an oxidized form.

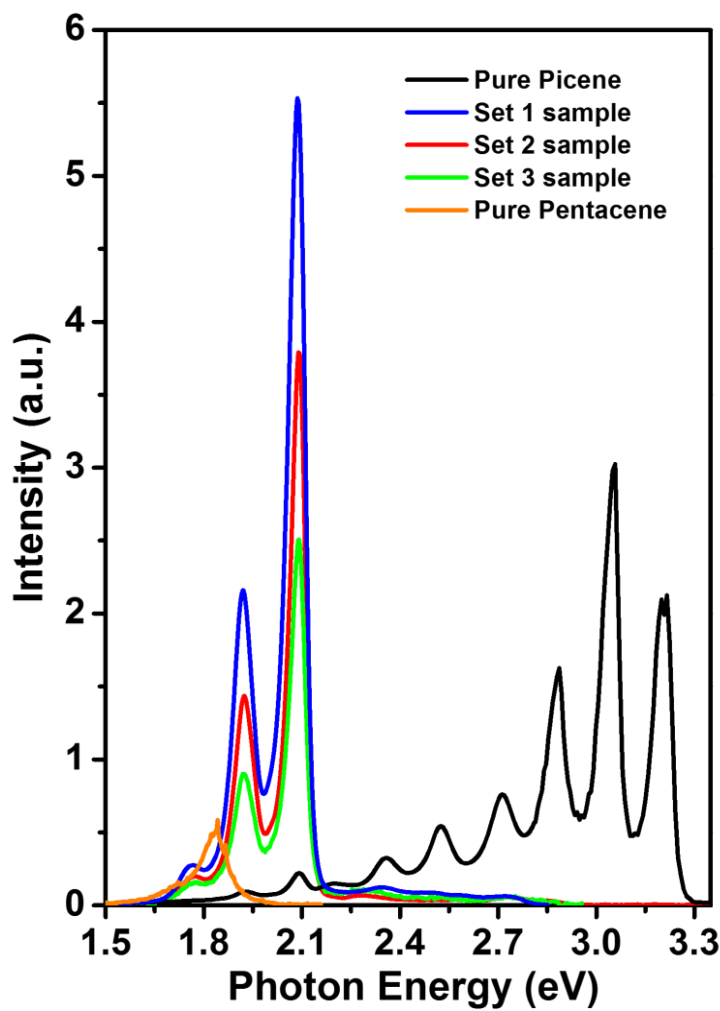

**Figure 5.** Photoluminescence spectra of the different samples excited at a wavelength of 300 nm. Black curve: pure picene; blue curve set 1 sample; red curve set 2 sample; green curve set 3 sample; Orange curve pure pentacene (in this case the excitation was made with a laser at 568 nm).

The main change in the photophysics of the doped samples becomes visible from the PL spectra. In Figure 5 we report the spectra of the doped samples and the PL of pure picene and pure pentacene for comparison. The black curve shows the emission spectra of picene,[36] with a structured emission band in the spectral range be-tween 3.4 eV and 2.7 eV characterized by the presence of peaks at 3.21 eV, 3.06 eV, 2.89 eV, 2.71 eV, due to the singlet-singlet transitions and the peaks at 2.53 eV, 2.38 eV, and 2.09 eV due to the triplet-singlet transition.[37] The extended and well-defined vibronic progression of this sample, underlines the high quality and crystallinity of the sample deposited by supersonic molecular beam.[22]

Therefore, the emission spectra relative to the samples doped with pentacene show completely different emission spectra and the spectra are also completely different respect to the typical emission of pure pentacene film.[37,38] We note an almost complete suppression of the picene emission for all sets of samples and at the same time the formation of a strong and sharp emission band at about 2.09 eV (with replicas at about 1.92 eV and 1.76 eV). We assign these bands to the emission coming from single pentacene molecules dispersed inside the picene matrix.[39] The absolute intensity of the emission that we observed for the samples doped with pentacene seems to be perceptibly larger than that of pure picene and we observe that the doped samples under UV illumination (i.e. mercury lamp) change their color from blue (as of pure picene) to orange. This observation is already an indication of the presence of energy transfer from the picene to the pentacene molecules that becomes clear from the time resolved measurements (see later). The fact that there is no red shift or change in line shape of the peaks increasing the pentacene concentration (effect visible in the absorbance in figure 4) indicates that for all samples the emission originates from the same excited state and thus only from molecules of pentacene well-dispersed inside the picene matrix. Pentacene aggregates, that may be present especially in the samples of set 3, might dissipate their excitation via non-radiative pathways thus decreasing their effective contribution to the intensity of the photoemission.

This assumption is supported by looking at the ratio between the picene PL peaks and pentacene PL peaks as a function of the pentacene concentration. After a progressive increase of intensity of the pentacene PL peak (at 2.09 eV) with increasing concentration of pentacene (i.e. observed at very low pentacene concentrations of less than one pentacene molecule every 105 molecules of picene) we observe a progressive decrease of the PL intensity with increasing the concentration of pentacene. We can rationalize this observation considering that where, up to a certain concentration of dopants, the energy transfer is limited by the amount of dopant molecules (i.e. dye saturation occurs). In practice, an optimal concentration of pentacene exists after which the distance between neighboring pentacene molecules is to close to avoid guest-guest interaction thus leading to a lower PL efficiency due to an increase of the non-radiative losses.

The progressive suppression of the picene emission and the enhancement of the pentacene optical features at the same time can be interpreted as a first indication of an efficient energy transfer between the picene and pentacene molecules.[30,31]

To shed light on the energy transfer between the picene matrix and the pentacene dopants we performed time resolved photoluminescence measurements. In Figure 6a the PL dynamics recorded at 2.75 eV (~450nm) for pure picene, set 2 samples and set 3 samples, are reported. At this energy we observe a remarkable reduction of the lifetime of the excited state of picene in the host guest system in comparison with the reference case of a pure picene film, reduction that is more pronounced for the sample of set 3. The PL of pure picene (black dots) at 445 nm displays a mono-exponential decay with time constant up to 10 ns. On the other hand, the set 2 sample displays a bi-exponential decay with time constants equal to 30 and 300 ps. Further reduction of the picene lifetime is demonstrated by samples with larger amount of pentacene dopant (set 3 sample). The strong quenching of the picene emission in presence of pentacene and its dependence respect to the pentacene concentration can be due to the resonant transfer of the excitation from picene to pentacene and to trapping of the photoexitation on the latter.

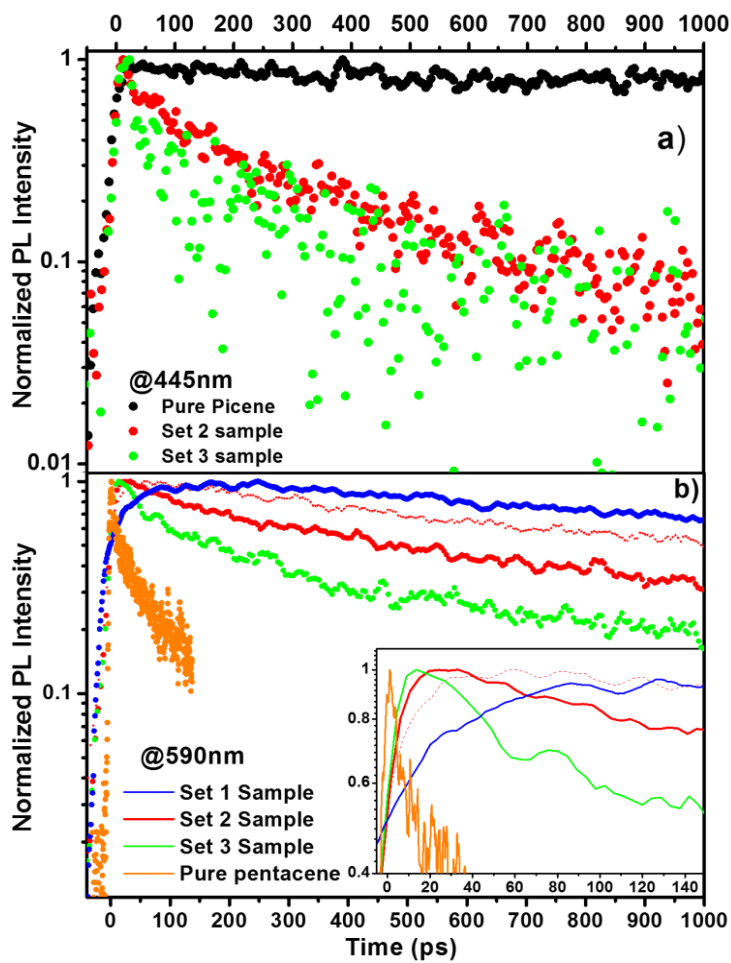

**Figure 6.** Photoluminescence dynamics of pentacene-doped picene thin films detected at 445nm (a) and at 590 nm (b). In figure 6b we report the dynamics of two samples of set 2 (in red), among the others, with slightly different pentacene concentrations with respect to 1 molecule of pentacene every $10^3$ picene molecules (Lower concentration [marked thin] and higher concentration [marked bold]. The inset in figure b) shows a close up view of the rise time of the different samples. The samples were excited at 380 nm.

To better understand these observation, in Figure 6b we report the time resolved photoluminescence decay of the samples of set 1, set 2, set 3 and pure pentacene for wavelengths corresponding to the main peak (0-0) of the pentacene optical emission (~ 2.09 eV). Looking in-to detail at the different dynamics we note that the samples of set 1 (lower pentacene concentration) show a mono-exponential behavior. This is maintained also for the sample of set 2 with the lower pentacene concentration (figure 6b). Here we found a lifetime of: $\tau \sim 1$ ns, which is typical of isolated molecules in a matrix. As the pentacene percentage increases the lifetime undergoes a reduction displaying a bi-exponential decay. The faster decay was shown for samples with higher pentacene concentration (set 3) where the τ being of 16 ps and 252 ps. It is important to emphasize that samples that display the longer lifetime are the ones that show the highest emission intensity. These observation suggest the presence of Förster energy transfer between the picene and pentacene molecules, which appears more efficient in samples having the lower pentacene concentration.

The bi-exponential behavior presented by the photo-luminescence dynamics of set 3 samples and of set 2 samples can be explained again with the formation of small clusters of pentacene, and with the opening of non-radiative channels due to the intermolecular interaction in the clusters. The formation of aggregates of pentacene in these samples is in agreement with what concluded from the analysis of the absorption spectra reported in Figure 4. Looking into detail the dynamics of the decay reported in figure 6b we observe the evidences for the samples at lower pentacene concentration of a rise time. In particular, we obtain a rise time up to 55 ps for the samples of set 1 at lower pentacene concentration. Shorter rise times were measured for samples with intermediate concentration (set 2 samples). Here we measure a rise time of 12 ps for the samples with a pentacene concentration lower than $10^{-3}$ molecules (red thin dashed line in the inset of figure 6b) and a rise time of 4 ps for the samples with pentacene concentration higher than $10^{-3}$ molecules (red bold line in figure 6b).

The presence of the rise time and its dependence respect to the pentacene concentration is an indication of an indirect excitation of the pentacene molecules. Förster energy transfer is the best candidate to explain such an indirect excitation due to the fast processes involved. The resonant energy level of picene excited state with the singlet of pentacene molecules and the overlap of the emission and absorption spectra of the host and guest respectively, suggest that a dominant contribution to the energy transfer could be of Förster type. At the same time the highly degree of crystallinity of these films increases the delocalization of the exciton favor in this way the energy transfer from the host to the guest molecules.

When the percentage of pentacene is increased, the rise time becomes faster. This can be due to the lower average distance between the photoexcited picene and the pentacene molecules; the direct photo-excitation of pentacene is less probable as the absorption of pentacene at the excitation wavelength is rather weak.[40] In fact, when we use an excitation energy lower than the picene optical gap, no luminescence is visible for all the set of samples reported.

The quenching of the picene emission in the doped samples indicates a high efficiency in the energy transfer process from picene to pentacene. The energy transfer give rise to an unusually strong PL coming from pentacene molecules. We believe that this is due to a breaking of the molecular symmetry of pentacene triggered by the crystalline field of picene. This system is thus very appealing for studying singlet fission in pentacene molecules and opens interesting possibilities for studying the photophysics of other excited states of pentacene (e.g. triplet state). The approach presented in this work may be extended also to other acenes.

## 4. Conclusion

In this work, we report a study of pentacene-doped picene thin films that were grown by supersonic molecular beam deposition. It is found that the pentacene-picene host-guest system can be an optimal system to become a reference in the study of doped nano-crystalline thin films made with small-molecules due to the complementary electronic and optical properties of these two molecules and the high similarities of the two stereoisomers. The photoluminescence spectra and time resolved measurements at room temperature and atmospheric pressure revealed an efficient energy transfer from the host matrix (picene) to the dopant molecules (pentacene) and high photo-stability of this host guest system. The electronic and excitonic energy levels of the two molecules favor an efficient energy transfer from the host to the guest. We attributed this energy transfer to a long range Förster resonant energy transfer, which is favored by the superposition of picene emission with pentacene absorption and the resonance of the emissive triplet state of picene and the singlet state of pentacene. Moreover, we explain the strong emission coming from pentacene (a molecule that usually shows a very low luminescence intensity) as due to its distortion in the crystalline field of picene. High population of singlets on the pentacene molecules enhance also the probability to have a transfer of the excitation, via inter system crossing, to the triplet state of pentacene which might be interesting for the study of solid state MASER based on pentacene.[16,17] From the results presented in this work, we also conclude that the wide optical-gap picene matrix could be a robust tool to investigate singlet fission in pentacene single molecules. The air stability and high mobility properties of picene, and the sharp emission peaks observed for this nanocrystal host-guest system paves the ways to its application as active material for light-emitting diodes, light-emitting transistors as well as light down-converter systems. Finally, the approach presented in this work may also be applied to other acenes and acene-derivatives molecules.

## Acknowledgements

The authors wish to thank Dr. A. Ghirri (CNR-Nano), R. Tat-ti (CNR-IMEM) and A. Verdini (CNR-IOM) for fruitfully discussion. Financial support from INFN-CNR national project (PREMIALE 2012) EOS